\documentclass{article}
\usepackage{arxiv}
\usepackage[utf8]{inputenc} 
\usepackage[T1]{fontenc}    
\usepackage{hyperref}       
\usepackage{url}            
\usepackage{booktabs}       
\usepackage{amsfonts}       
\usepackage{nicefrac}       
\usepackage{microtype}      
\usepackage{lipsum}	
\usepackage{amsmath}
\usepackage{tikz}
\usetikzlibrary{positioning, arrows.meta, shapes.geometric}
\usepackage{algorithm}
\usepackage{algpseudocode}
\usetikzlibrary{positioning, arrows.meta, shadows, shapes.geometric}
\usepackage{graphicx}
\usepackage[numbers]{natbib}
\usepackage{doi}
\usepackage{booktabs}
\usepackage{tcolorbox}
\usepackage{geometry}
\usepackage{indentfirst}
\usepackage{enumitem}
\usepackage{float}
\graphicspath{ {./images/} }
\title{ \textsc{\large Hybrid Vector Auto Regression and Neural Network Model for Order Flow Imbalance Prediction in High‑Frequency
Trading}}

\author{ \href{https://sites.google.com/view/shafiq-abdulrahman-iitm/home}{\hspace{1mm}Abdul Rahman}{} \\
	Department of Mathematics\\
	Indian Institute of Technology,Madras\\
	Chennai,India \\
	\texttt{abdul.math@alumni.iitm.ac.in}
   \And
 \href{https://math.iitm.ac.in/neelesh}{Neelesh Upadhye}\\
  Department of Mathematics\\
	Indian Institute of Technology,Madras\\
	Chennai,India \\
	\texttt{neelesh@iitm.ac.in}
}

\begin{document}
\maketitle
\begin{abstract}
\hspace{0.2cm} In high-frequency trading, accurate prediction of Order Flow Imbalance (OFI) is crucial for understanding market dynamics and maintaining liquidity. This paper introduces a hybrid predictive model that combines Vector Auto Regression (VAR) with a simple feedforward neural network (FNN) to forecast OFI and assess trading intensity. The VAR component captures linear dependencies, while residuals are fed into the FNN to model non-linear patterns, enabling a comprehensive approach to OFI prediction. Additionally, the model calculates the intensity on the Buy or Sell side, providing insights into which side holds greater trading pressure. These insights facilitate the development of trading strategies by identifying periods of high buy or sell intensity. Using both synthetic and real trading data from Binance, we demonstrate that the hybrid model offers significant improvements in predictive accuracy and enhances strategic decision-making based on OFI dynamics. Furthermore, we compare the hybrid model’s performance with standalone FNN and VAR models, showing that the hybrid approach achieves superior forecasting accuracy across both synthetic and real datasets, making it the most effective model for OFI prediction in high-frequency trading.

\end{abstract}


\section{Introduction}

Order Flow Imbalance (OFI) is a critical indicator in high-frequency trading (HFT) and financial markets more broadly, offering insights into market sentiment and directional pressure by capturing the net difference between buy and sell orders. In the fast-paced realm of HFT, understanding OFI allows market participants to assess imminent price fluctuations based on the real-time distribution of orders within the limit order book. The imbalance between buy and sell orders, measured over very short intervals, serves as an indicator of underlying supply-demand conditions, often preceding price movements. Studies such as those by  \citet{Cont2023Cross-impactMarkets} underscore the role of OFI as a driver of short-term price changes, linking imbalances to price impact and liquidity shifts. The ability to accurately model and forecast OFI, therefore, holds significant implications for HFT strategies and liquidity management.

The mechanism of OFI is rooted in the interactions of market participants—liquidity providers, arbitrageurs, institutional traders, and market makers—who bring diverse motivations to the order book. As market participants place buy or sell orders, the order book’s dynamics shift continuously, often leading to imbalances that can influence asset prices. Traditional models in market microstructure, such as those discussed by  \citet{Easley2012FlowWorld}, have highlighted the relationship between order flow and liquidity, with OFI acting as a real-time indicator of market conditions. When buy orders dominate, a positive OFI indicates potential upward pressure on price, whereas a negative OFI, caused by an excess of sell orders, signals likely downward movement. The sensitivity of OFI to order book shifts makes it a useful tool for identifying trends and gauging momentum, which can assist traders in optimizing execution and minimizing market impact. \citet{Kolm2023DeepBook} leverages Order Flow Imbalance (OFI) data from the limit order book to train deep learning models that predict price movements across multiple trading horizons, helping traders capture alpha, or abnormal returns, through short-term price forecasting. The approach demonstrates that OFI-based models are robust and effective for high-frequency trading strategies by identifying predictive signals that outperform traditional methods.

Despite its importance, modeling OFI remains challenging due to the noisy and volatile nature of high-frequency data. Econometric time series models such as Vector Auto Regression (VAR) have been widely used to capture linear dependencies in financial data. VAR models provide interpretability and are particularly effective in examining interdependencies among multiple time series, as highlighted in studies on credit and liquidity factors by \citet{Murphy2012AMarkets}. However, VAR's linear framework limits its applicability in high-frequency contexts where price dynamics are often non-linear and can shift abruptly. As noted by \citet{Bacry2016EstimationDynamics}, high-frequency trading data contains complex patterns driven by market microstructure that cannot always be captured with linear models alone.

Machine learning, particularly neural networks, has emerged as a promising approach to overcome the limitations of linear models in capturing non-linear dynamics. Neural networks, with their ability to model intricate patterns and adapt to non-linear relationships, have been applied in financial time series forecasting with notable success. Studies by \citet{HUANG2007NEURALFORECASTING} demonstrate that neural networks can effectively capture the non-linear dependencies present in high-frequency data, improving accuracy in predictions of financial metrics such as volatility and return. However, neural networks come with limitations, including potential overfitting, high computational costs, and the requirement for large training datasets to generalize effectively, particularly when dealing with volatile high-frequency trading data.

To address these limitations, hybrid models combining the strengths of both econometric time series and machine learning techniques have been proposed. These hybrid approaches, such as the one employed by \citet{Maleki2020AsymmetricData} for analyzing heavy-tailed vector auto-regressive processes, enable the capturing of linear dependencies while also accounting for non-linear residuals . The integration of traditional VAR models with feedforward neural networks (FNNs) is motivated by Universal Differential Equations (UDE), which leverage known dynamics while allowing neural networks to learn residual patterns. This hybrid approach has been shown to enhance model flexibility and improve predictive performance in complex, non-linear environments . By using VAR to capture linear patterns and FNNs to model non-linear residuals, the hybrid model can more accurately capture OFI dynamics, particularly in high-frequency trading contexts where both linear and non-linear influences coexist.

In this study, we present a hybrid VAR-FNN model designed for OFI prediction in high-frequency trading. By integrating a linear VAR component with a non-linear neural network layer, this model aims to address the unique challenges presented by high-frequency data, leveraging the interpretability of VAR with the adaptability of neural networks. Furthermore, we introduce an additional intensity metric, representing the magnitude of directional trading pressure on the Buy and Sell sides, to enhance strategic decision-making. Through experiments on both synthetic and real-world datasets, we demonstrate that the proposed model not only achieves higher accuracy than standalone models but also provides valuable insights for trading strategies. The combination of VAR and FNN creates a robust framework for OFI prediction that aligns with the demands of high-frequency trading, offering potential applications for liquidity management and optimal trade execution.

\section{Literature Review and Related Works}

Order Flow Imbalance (OFI) serves as a critical metric for understanding market dynamics in high-frequency trading (HFT). It captures the net difference between buy and sell orders, providing insights into directional market pressure and anticipating price fluctuations. The relevance of OFI has been well-established in financial literatures like \citet{Cont2023Cross-impactMarkets} , \citet{Smales2013BondImbalance},  \citet{Chan2000TradeRelation} with studies consistently showing its influence on price impact, trade size,  liquidity, and volatility . For example, \citet{Cont2014TheEvents} discuss the price impact of order book events and demonstrate that OFI can effectively predict short-term price movements . However, modeling OFI remains challenging due to the noisy and volatile nature of high-frequency trading data, which often necessitates advanced approaches to capture both linear and non-linear dependencies. In high-frequency trading, we classify market orders as ‘BUY’ or ‘SELL’ . The classification of each trade as ‘BUY’ or ‘SELL’ can be deterministically inferred by tracing the origin of the passive order.

A formal definition of OFI is given as follows:

\begin{equation}
\text{OFI}(T, h) = \frac{\Delta N^{B}_{T - h, T} - \Delta N^{S}_{T - h, T}}{\Delta N^{B}_{T - h, T} + \Delta N^{S}_{T - h, T}}
\end{equation}

where OFI is measurable at time \( T \) over a window length \( h \) and is defined by the following parameters:
\begin{enumerate}
    \item \( T \in \{ t_1 = 0, \dots, t_i, \dots, t_N = T_F \} \), where \( t_i - t_{i-1} = 1 \) for all \( i \).
    \item \( h \): Window length over which trades are counted.
    \item \( \Delta N^{S}_{T - h, T} = N^{S}_T - N^{S}_{T - h} \): Number of ‘SELL’ (S) classified trades over the window \( h \), with \( T \) as the present time.
    \item \( \Delta N^{B}_{T - h, T} = N^{B}_T - N^{B}_{T - h} \): Number of ‘BUY’ (B) classified trades over the window \( h \), with \( T \) as the present time.
\end{enumerate}

To calculate the trading intensity signal \( \sigma \) based on the Order Flow Imbalance (OFI), we follow these steps:

\begin{itemize}
    \item \textbf{Order Flow Imbalance Range:} The OFI is bounded by \( -1 \leq \text{OFI} \leq 1 \), where values close to 1 indicate strong buying pressure, and values close to -1 indicate strong selling pressure.
    
    \item \textbf{Threshold Definition:} We define a positive threshold \( T \) (where \( 0 \leq T < 1 \)), typically close to zero, to filter out noise and identify significant buy or sell signals.

    \item \textbf{Signal Calculation:} The trading signal \( \sigma \) is calculated as follows:
    \[
    \sigma = 
    \begin{cases} 
      \text{"BUY"} & \text{if } \text{OFI} > T, \\
      \text{"SELL"} & \text{if } \text{OFI} < -T, \\
      \text{"HOLD"} & \text{if } -T \leq \text{OFI} \leq T.
    \end{cases}
    \]
\end{itemize}

Thus:
\begin{itemize}
    \item If \( \text{OFI} > T \), the signal \( \sigma \) is set to "BUY," indicating an upward trend or buying pressure.
    \item If \( \text{OFI} < -T \), the signal \( \sigma \) is set to "SELL," indicating a downward trend or selling pressure.
    \item If \( -T \leq \text{OFI} \leq T \), the signal \( \sigma \) is set to "HOLD," suggesting market neutrality or a balanced state.
\end{itemize}

This setup ensures that signals are generated only when OFI values exceed the threshold \( T \), reducing noise and capturing meaningful trends.

\begin{table}[h!]
\centering
\caption{\small Sample of Level 1 Limit Order Book Data with Calculated Order Flow Imbalance and Trading Intensity}
\label{tab:order_book_sample}
\small
\begin{tabular}{@{}p{0.8cm} p{1cm} p{1cm} p{1cm} p{1cm} p{1.5cm} p{1.2cm}@{}}
\toprule
Time & Best Bid & Best Ask & Buy \hspace{0.2cm}Orders & Sell \hspace{0.2cm}Orders & Order Flow Imbalance & Trading Intensity \\ \midrule
T1   & 100.5 & 101.0 & 55 & 30 & 0.294 & BUY \\
T2   & 100.6 & 101.2 & 45 & 40 & 0.059 & HOLD \\
T3   & 100.7 & 101.3 & 60 & 125 & -0.351 & SELL \\ \bottomrule
\end{tabular}
\end{table}

\subsection{Vector Auto Regression (VAR) in Financial Modeling}

The Vector Auto Regression (VAR) model \cite{Stock2001VectorAutoregressions, Toda1994VectorStudy}  is a powerful tool in financial time series analysis, widely valued for its ability to capture linear interdependencies among multiple variables in a system . In a VAR model, each variable within a multivariate time series is predicted based on its past values as well as the historical values of other time series within the system, which allows for a comprehensive analysis of mutual influences. This capability is especially advantageous in finance, where understanding the interconnected behaviors of indicators like interest rates, exchange rates, and asset prices is essential for analyzing systemic dependencies . \citet{Maleki2020AsymmetricData} demonstrated the efficacy of such models in capturing diverse dependencies in financial time series . By modeling these interactions, VAR provides insights into the broader market dynamics, making it a foundational approach in econometric forecasting and financial risk assessment.

To explore the behavior of OFI, \citet{Anantha2024ForecastingImbalance} constructed a time series comprising minute-level OFI data for a single trading day, resulting in 375 observations. Through visual inspection using autocorrelation function (ACF) and partial autocorrelation function (PACF) plots, they identified significant autocorrelation in OFI, especially at recent time points. This finding motivated the use of the VAR model as a suitable approach for capturing OFI's dependency on recent history. The observed autocorrelation characteristics underscore the need for advanced time series models, such as VAR, in high-frequency trading, where OFI dynamics exhibit both linear and non-linear patterns.

A VAR model of lag order \( p \) is represented as follows:

\begin{equation}
\mathbf{Y}_t = \mathbf{c} + \sum_{i=1}^{p} \mathbf{A}_i \mathbf{Y}_{t-i} + \mathbf{\varepsilon}_t,
\end{equation}

where:
\begin{itemize}
    \item \( \mathbf{Y}_t \) - is a  \( k \times 1 \) vector of time series variables at time \( t \), where \( k \) denotes the number of variables in the system.
    \item \( \mathbf{c} \) -is a  \( k \times 1 \) vector of constants, representing the intercept terms for each time series.
    \item \( \mathbf{A}_i \) -is a   \( k \times k \) matrix of coefficients for each lag \( i \), which captures the linear relationship between the time series variables at time \( t \) and their values at time \( t - i \).
    \item \( \mathbf{\varepsilon}_t \) - is a  \( k \times 1 \) vector of white noise errors at time \( t \), assumed to have zero mean and constant variance, \( \mathbf{\varepsilon}_t \sim \mathcal{N}(0, \Sigma) \).
\end{itemize}

In this framework, each element of \( \mathbf{Y}_t \) depends on the past \( p \) values of all \( k \) variables, allowing the model to account for mutual dependencies across multiple time series. The choice of lag \( p \) is critical, as it determines the extent to which past observations are considered, and is often determined through criteria such as the Akaike Information Criterion (AIC) or Bayesian Information Criterion (BIC).

The VAR model has been extensively applied in financial contexts, particularly for capturing interdependencies in macroeconomic and market data. \citet{Murphy2012AMarkets} applied VAR to analyze credit and liquidity factors in LIBOR and Euribor swap markets, demonstrating its efficacy in modeling dynamic interrelationships in financial systems . However, one limitation of the VAR model is its assumption of linearity, which may restrict its applicability in high-frequency financial data characterized by more complex, non-linear dependencies. For example, high-frequency trading data often contains intricate patterns driven by market microstructure effects, which cannot be fully captured by linear models .

In this study, the VAR model serves as the initial component in a hybrid approach to modeling OFI. By leveraging VAR, we aim to capture the linear dependencies present in high-frequency OFI data, thereby providing a foundational layer of interpretability. The residuals from this linear model are then fed into a feedforward neural network (FNN) to account for the remaining non-linear dependencies, creating a comprehensive model that addresses both the linear and non-linear aspects of OFI dynamics.This study draws on the significant autocorrelation observed in \citet{Anantha2024ForecastingImbalance} analysis of OFI dynamics  While Anantha and Jain applied a Hawkes process for OFI modeling, our hybrid approach leverages this autocorrelation insight to inform the use of a VAR model. Combining the interpretability of VAR with the flexibility of neural networks, our hybrid framework offers a robust solution for predicting OFI in high-frequency trading contexts

\subsection{Machine Learning and Neural Networks in Financial Modeling}

Machine learning has become transformative in financial modeling, enabling the discovery and analysis of complex, non-linear patterns that traditional econometric time series models often miss. Studies by \citet{Burrell1997TheFinance} demonstrate the impact of neural networks in finance. Feedforward neural networks (FNNs), in particular, have become prominent in financial applications due to their adaptability and capacity to model intricate dependencies within data.  This adaptability is crucial in finance, where non-linear relationships frequently emerge from market sentiment, investor behavior, and macroeconomic trends . FNNs have proven effective in predicting essential metrics, such as asset returns and volatility, which are influenced by these multifaceted factors.

Moreover, the scalability and flexibility of neural networks \cite{Vellido1999Neural19921998} allow them to handle the high-frequency and high-volume data characteristic of modern financial markets. This capability enables FNNs to adapt effectively to volatile environments, providing insights for both strategic forecasting and short-term trading. As financial data becomes increasingly complex, neural networks offer robust tools for real-time analysis \cite{Burrell1997TheFinance}, outperforming traditional models constrained by linear assumptions .

Feedforward neural networks operate by passing inputs through layers of interconnected neurons, each with adjustable weights, to capture non-linear mappings from inputs to outputs. Studies by \citet{HUANG2007NEURALFORECASTING} demonstrate the efficacy of FNNs in financial forecasting tasks, such as predicting volatility and returns . In these applications, FNNs can identify underlying patterns that linear models may overlook, making them a valuable tool for complex financial time series analysis. Despite their potential, however, FNNs face limitations when applied to high-frequency trading data, where the risk of overfitting is higher due to the inherently noisy nature of the data. Moreover, training neural networks on large, high-frequency datasets can be computationally intensive, requiring significant processing power and memory, particularly for deep architectures.

To address these challenges, hybrid models that combine the strengths of traditional econometric time series methods and neural networks have been proposed. Such models leverage the interpretability of econometric time series models, like Vector Auto Regression (VAR), while capturing non-linear residuals through neural networks. For example, \citet{Maleki2020AsymmetricData}  applied a similar approach to financial time series with heavy-tailed distributions, demonstrating how a hybrid model could achieve more accurate predictions than standalone models . In our hybrid approach, we utilize a VAR model to capture the linear dependencies within Order Flow Imbalance (OFI) data. The residuals from the VAR model, which may contain non-linear patterns, are then fed into an FNN. This layered approach enables the model to capture both linear dependencies and complex, non-linear residuals within the data.

While standalone neural networks can perform well on financial data, the hybrid model provides an incremental improvement in accuracy, especially when applied to high-frequency datasets. By first using a VAR model to account for linear relationships, we reduce the complexity of the non-linear patterns that the FNN must capture, mitigating the risks of overfitting and high computational demands. On our datasets, the hybrid model consistently yields slightly better performance metrics than either the standalone VAR or FNN models, confirming the benefit of combining these methods.

\begin{figure}[h!]
\centering
    \includegraphics[width=0.7\textwidth]{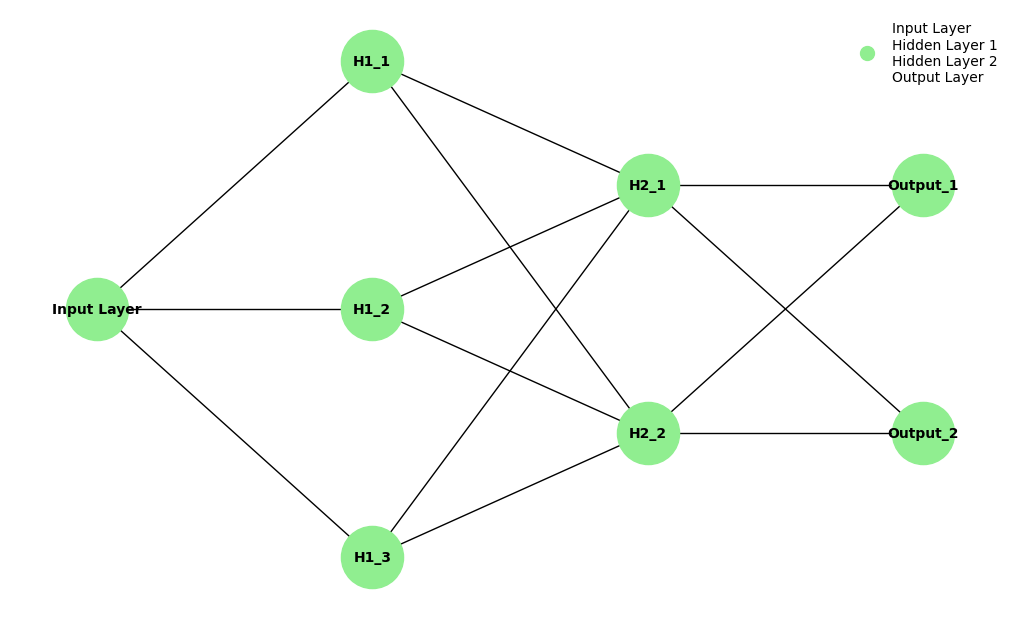}
\caption{Basic architecture of a feedforward neural network (FNN) for financial prediction.}
\label{fig:fnn_architecture}
\end{figure}

The basic architecture of the FNN used in our model is shown in Figure \ref{fig:fnn_architecture}. This architecture consists of an input layer corresponding to the residuals from the VAR model, multiple hidden layers for learning non-linear patterns, and an output layer that produces OFI predictions. Each neuron within the hidden layers applies a non-linear activation function, which allows the network to model the complex patterns present in high-frequency financial data.

Overall, the use of a hybrid VAR-FNN model aligns with advancements in financial machine learning, where integrating traditional and modern techniques helps to balance interpretability and accuracy. While neural networks offer powerful tools for modeling non-linear dependencies, combining them with established econometric time series models provides a comprehensive approach for high-frequency trading analysis, enhancing the robustness and accuracy of OFI predictions.

\subsection{Hybrid Modeling Approaches: VAR and Neural Network Integration}

Hybrid models combining Vector Auto Regression (VAR) with neural networks present a balanced approach for modeling Order Flow Imbalance (OFI) in high-frequency trading. These models capitalize on the interpretability of econometric time series methods and the adaptability of machine learning, making them well-suited for complex datasets with both linear and non-linear dependencies.

In a hybrid VAR-FNN model, the VAR component captures linear dependencies, while the feedforward neural network (FNN) models non-linear residuals, which are not accounted for by VAR. This approach aligns with the Universal Differential Equations (UDE) framework, which combines known dynamics with neural networks to learn unknown patterns . Mathematically, this can be represented as:

\begin{equation}
\text{OFI}_t = f(\text{Residual}_t),
\end{equation}

where \( f \) is the non-linear function learned by the FNN on VAR residuals, capturing complexities beyond the scope of linear models.

Hybrid models have shown improved accuracy in applications like trading strategies and volatility forecasting. For instance, \citet{Cartea2018EnhancingSignals} found that hybrid models enhance trading strategies by providing more reliable short-term predictions . By leveraging both linear and non-linear modeling, the hybrid VAR-FNN model offers a comprehensive approach to OFI prediction, yielding incremental gains over standalone models.

\begin{table}[h!]
\centering
\caption{ \small{Comparison of Models for Order Flow Imbalance (OFI) Prediction}}
\label{tab:model_comparison}
\resizebox{\columnwidth}{!}{%
\begin{tabular}{@{}lcc@{}}
\toprule
Model Type      & Strengths                              & Limitations                      \\ \midrule
VAR             & Captures linear dependencies; interpretable & Ineffective for non-linear patterns \\
FNN (Neural Network) & Models non-linear dependencies; flexible adaptation & Risk of overfitting; high computational cost \\
Hybrid VAR-FNN  & Combines linear and non-linear modeling; improved accuracy & Increased model complexity; higher training time \\ \bottomrule
\end{tabular}%
}
\end{table}

\subsection{Alternative Modeling Approaches for OFI Forecasting}

In addition to VAR, FNN, and hybrid models, a range of other methodologies has been employed to forecast OFI in high-frequency trading, each providing unique insights into market behavior. One notable approach is the Hawkes Process model, which captures the self-exciting nature of order arrivals and effectively models the temporal clustering of buy and sell orders. \citet{Anantha2024ForecastingImbalance} demonstrated the efficacy of the Hawkes Process in high-frequency trading environments by forecasting OFI with improved short-term accuracy, highlighting its robustness in handling the high interdependence of events over time . However, studies from \citet{Bacry2014HawkesDynamics} results that Hawkes processes can accurately model the cascading effects within high-frequency trading, they are computationally intensive, especially for real-time applications.

Another approach is the Hidden Markov Model (HMM), which forecasts OFI by identifying latent market states to anticipate directional market shifts. \citet{Wu2020CapturingKOSPI50} used HMMs to model order imbalance in the SET50 and KOSPI50 markets, identifying market states characterized by increased buy or sell pressure. This approach proved effective for classifying market phases and tracking directional changes in market sentiment . Although HMMs are valuable for recognizing market states, their practical application in high-frequency trading is often restricted due to the complexity and computational cost of processing high-dimensional data in real-time.

Dynamic modeling of OFI has also been approached through optimal execution models, which use order imbalance to gauge market liquidity . \citet{Bechler2015OptimalImbalance} applied dynamic order flow imbalance models for optimal execution, demonstrating that OFI can be a reliable predictor for balancing trade execution while minimizing market impact . This strategy has shown effectiveness in ensuring favorable execution while accounting for liquidity constraints. However, these models typically focus on execution strategies rather than the broader scope of OFI forecasting across various trading intensities.

Studies on OFI have also leveraged models examining the relationship between order flow imbalance and market efficiency. For example, \citet{Jiang2011OrderMarket} explored how order imbalance impacts liquidity and market efficiency in the Chinese stock market, showing that OFI strongly correlates with liquidity levels and affects price discovery in emerging markets . This research underscores the impact of OFI on liquidity but focuses primarily on market-wide dynamics rather than high-frequency prediction of OFI.

Gradient Boosting Machines, particularly models like XGBoost, have been successfully employed to capture non-linear relationships between order flow and price impact. \citet{Cartea2018EnhancingSignals} used gradient boosting to predict OFI, achieving high accuracy in modeling the dependencies between trades and market order imbalances . Gradient boosting methods, while effective, tend to require careful regularization to avoid overfitting in noisy, high-frequency data environments.

Bayesian Networks have also been employed to model OFI probabilistically, estimating the likelihood of directional shifts in the market. \citet{Easley2012FlowWorld} explored Bayesian methods for modeling order flow toxicity, with implications for predicting OFI, finding that Bayesian networks could provide reliable probabilistic estimates of buy or sell dominance. Despite the interpretability and accuracy of Bayesian networks, they can be computationally expensive, making them challenging to implement in high-frequency contexts.

These alternative approaches highlight the diverse methodologies available for OFI forecasting, each with specific advantages and limitations. While models such as Hawkes processes and HMMs excel in capturing event-driven dynamics and market state transitions, execution models and Bayesian networks provide insights into liquidity and directional probabilities. The choice of model, therefore, depends on the desired trade-off between accuracy, interpretability, and computational efficiency, particularly in high-frequency trading settings.

\subsection{Motivation for Hybrid Modeling and Research Gaps}
 The alternative models provided in previous sections offer unique strengths and weaknesses, with accuracy levels typically ranging from moderate to high, depending on the data's time resolution and market volatility. Hybrid models, which integrate both traditional and machine learning approaches, are increasingly recognized as effective solutions for capturing the multifaceted patterns in OFI data, particularly in high-frequency contexts where both linear and non-linear dynamics are present.
Accurate modeling of Order Flow Imbalance (OFI) is crucial for developing trading strategies and managing liquidity in high-frequency markets \cite{Fung2007OrderPrices,Kromkowski2016EffectStates}. The ability to predict directional order flow allows market participants to anticipate price movements and adjust their trading behavior accordingly. Studies by \citet{Smales2013BondImbalance} and \citet{Su2012DynamicGainers}  underscore the relationship between OFI and price impact, noting that order imbalance can significantly affect asset prices and volatility . This study’s focus on combining VAR with FNN for OFI prediction aligns with the broader trend in financial research to integrate multiple modeling techniques for enhanced predictive power .

Despite advancements in hybrid modeling, research specifically addressing hybrid VAR-FNN models for OFI prediction in high-frequency trading remains limited. Most studies have explored VAR or neural network models independently, missing the potential benefits of integrating both approaches. This study addresses this gap by developing a VAR-FNN hybrid model tailored for OFI prediction, incorporating an intensity metric to provide insights into directional market pressure. The model is evaluated on both synthetic and real datasets to assess its robustness and generalizability, contributing a novel, integrated econometric time series models and machine learning approach to OFI prediction.

\section{Methodology}

This section details the steps taken in the modeling process, including data preparation, FNN modeling, and the hybrid VAR-FNN model. The algorithms provided here highlight the process for training the FNN model and Hybrid VAR -FNN models, as well as the workflow for the hybrid model.
\\

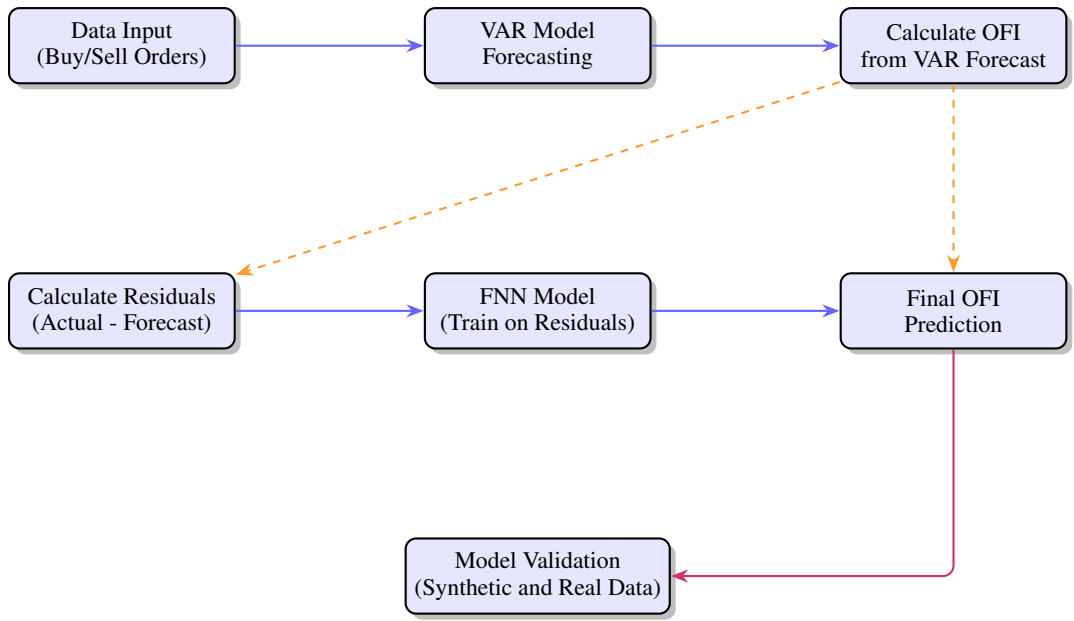
\begin{figure}[h!]
\centering
\begin{tikzpicture}[
    node distance=1.5cm and 2.5cm,
    every node/.style={rectangle, rounded corners, draw=black, thick, align=center, font=\small, minimum width=3cm, minimum height=1cm, fill=blue!10, drop shadow},
    arrow/.style={-Stealth, thick, draw=blue!60, rounded corners},
    arrow2/.style={-Stealth, thick, draw=orange!80, dashed, rounded corners},
    arrow3/.style={-Stealth, thick, draw=purple!80, rounded corners, bend right=45}
]

\node (input) {Data Input \\ (Buy/Sell Orders)};
\node (var) [right=of input] {VAR Model \\ Forecasting};
\node (ofi_calc) [right=of var] {Calculate OFI \\ from VAR Forecast};

\node (residuals) [below=of input, yshift=-1cm] {Calculate Residuals \\ (Actual - Forecast)};
\node (fnn) [right=of residuals] {FNN Model \\ (Train on Residuals)};
\node (ofi_pred) [right=of fnn] {Final OFI \\ Prediction};

\node (validate) [below=of fnn, yshift=-1cm] {Model Validation \\ (Synthetic and Real Data)};

\draw[arrow] (input) -- (var);
\draw[arrow] (var) -- (ofi_calc);
\draw[arrow2] (ofi_calc) -- (residuals);
\draw[arrow2] (ofi_calc) -- (ofi_pred);
\draw[arrow] (residuals) -- (fnn);
\draw[arrow] (fnn) -- (ofi_pred);
\draw[arrow3] (ofi_pred) |- (validate);


\end{tikzpicture}
\caption{Workflow of the Hybrid VAR-FNN Model for OFI Prediction}
\label{fig:hybrid_model_flow}
\end{figure}

\subsection{Process Overview}

The methodology follows a structured workflow:
\begin{itemize}
    \item \textbf{Data Preparation}: Input data includes synthetic and real-world datasets, specifically buy and sell orders.
    \item \textbf{VAR Forecasting}: The VAR model is used to forecast future buy and sell orders, from which initial OFI values are calculated.
    \item \textbf{Residual Calculation}: Residuals between actual buy/sell orders and VAR forecasts are computed to capture non-linear patterns.
    \item \textbf{FNN Training on Residuals}: The FNN is trained on these residuals to learn the non-linear patterns not captured by the VAR model.
    \item \textbf{Final OFI Prediction}: The final OFI values are obtained by combining the VAR forecasted OFI and FNN-predicted residuals.
    \item \textbf{Model Validation}: Validation on synthetic and real datasets to ensure robustness.
\end{itemize}


\subsection{Algorithm for FNN Modeling}

The FNN-only modeling algorithm is used to predict OFI by directly learning from input data of buy and sell orders.

\begin{algorithm}
\caption{FNN Modeling for OFI Prediction}
\label{algo:fnn}
\begin{algorithmic}[1]
\Require Initial parameters $\Theta^{(0)}$, number of epochs $M$
\Ensure Final parameters $\Theta^{(M)}$
\State \textbf{Input:} Buy and Sell Orders
\State \textbf{Output:} Predicted OFI
\For{$i = 0$ to $M - 1$}
    \State Compute $\nabla L(\Theta^{(i)})$ for loss function $L$
    \State Update $\Delta \Theta^{(i)} = - \nabla L(\Theta^{(i)})$
    \State Select learning rate $\gamma$
    \State Update parameters $\Theta^{(i+1)} = \Theta^{(i)} + \gamma \cdot \Delta \Theta^{(i)}$
\EndFor
\State \Return Final parameters $\Theta^{(M)}$ and OFI predictions
\end{algorithmic}
\end{algorithm}

\textbf{Time Complexity:} The time complexity of this FNN-only modeling algorithm is approximately:
\[
\mathcal{O}(M \cdot n \cdot d \cdot h)
\]
where \( M \) is the number of epochs, \( n \) is the number of data points, \( d \) is the dimensionality of the input data, and \( h \) is the number of neurons in the hidden layers.
(see Appendix~\ref{appendix:A2})

\subsection{Algorithm for Hybrid VAR-FNN Model}

The hybrid model workflow leverages both VAR and FNN to capture linear and non-linear patterns in OFI prediction. The VAR component forecasts buy and sell orders, while the FNN model learns from the residuals.

\begin{algorithm}
\caption{Hybrid VAR-FNN Modeling for OFI Prediction}
\label{algo:hybrid_var_fnn}
\begin{algorithmic}[1]
\Require Initial parameters $\Theta_{\text{FNN}}^{(0)}$, number of epochs $M$, lag order $p$ for VAR model
\Ensure Final OFI prediction
\State \textbf{Input:} Buy and Sell Orders
\State \textbf{Output:} Final OFI Prediction
\State Train VAR model on buy/sell orders to forecast future orders
\State Calculate initial OFI values from VAR forecasts
\State Compute residuals: $\text{Residuals} = \text{Actual Orders} - \text{VAR Forecasted Orders}$
\For{$i = 0$ to $M - 1$}
    \State Train FNN on residuals to capture non-linear patterns
    \State Update $\Delta \Theta_{\text{FNN}}^{(i)} = - \nabla L(\Theta_{\text{FNN}}^{(i)})$
    \State Update parameters $\Theta_{\text{FNN}}^{(i+1)} = \Theta_{\text{FNN}}^{(i)} + \gamma \cdot \Delta \Theta_{\text{FNN}}^{(i)}$
\EndFor
\State Combine VAR forecasted OFI and FNN-predicted residuals to obtain final OFI predictions
\State \Return Final OFI predictions for model validation
\end{algorithmic}
\end{algorithm}

\textbf{Time Complexity:} The overall time complexity of this hybrid VAR-FNN model is approximately $\mathcal{O}(n \cdot p^2) + \mathcal{O}(M \cdot n \cdot d \cdot h)$, where $n$ is the number of data points, $p$ is the VAR lag order, $M$ is the number of epochs, $d$ is the input dimensionality, and $h$ represents the number of neurons in the hidden layers of the FNN.
(see Appendix~\ref{appendix:A3})

\subsection{Evaluation Metrics}

The model performance is evaluated using the following metrics:
\begin{itemize}
    \item \textbf{Mean Absolute Error (MAE)}: Measures the average error magnitude in OFI predictions.
    \item \textbf{Mean Squared Error (MSE)}: Provides a squared measure of error, penalizing larger errors.
    \item \textbf{R-Squared ($R^2$)}: Indicates the proportion of variance explained by the model.
    \item \textbf{Accuracy and Precision on Trading Intensity Signals}: Evaluates the model’s effectiveness in predicting directional trading pressures -(BUY or SELL).
\end{itemize}

These metrics enable a comparative analysis of the FNN and hybrid VAR-FNN models, providing insight into each model’s accuracy and robustness.








\section{Experiments and Results}

This section provides a comprehensive overview of the data used, model configurations, sensitivity analysis of key parameters, and the resulting performance of both the standalone Feedforward Neural Network (FNN) and hybrid VAR-FNN models. The models were trained and validated on high-frequency cryptocurrency data sourced from Binance.

\subsection{Data Details}

To develop and validate the models, we utilized a dataset of \textbf{10,000 seconds} of continuous buy and sell order data, gathered for a single cryptocurrency without applying any normalization or resampling. For model validation, we employed three datasets to assess the model’s robustness across different scenarios:
\begin{itemize}
    \item \textbf{Real Dataset 1 (BTCUSD):} Contains approximately 3,000 data points representing BTCUSD transactions.
    \item \textbf{Real Dataset 2 (ETCUSDT):} Contains approximately 3,000 data points for ETCUSDT transactions.
    \item \textbf{Synthetic Dataset:} A synthetically generated dataset of 3,000 data points designed to replicate characteristics similar to the real datasets.
\end{itemize}

\subsection{Sensitivity Analysis and Optimal Parameter Selection}

\subsubsection{Parameter Configuration}
To optimize the performance of the hybrid VAR-FNN model, we conducted a sensitivity analysis on several key parameters:
\begin{itemize}
    \item \textbf{Lag Order (p) in VAR:} Examined values were 1, 2, 5, and 10, determining the extent of temporal dependencies captured by the VAR component.
    \item \textbf{FNN Architecture:} Different layer configurations were tested, including 128-64-2, 32-16-2, 32-32-2, 128-64-32-2, and 64-32-16-2 neurons per layer.
    \item \textbf{Activation Functions:} ReLU, Tanh, and Sigmoid activations were applied to observe their impact on the model’s non-linear pattern recognition capabilities.
    \item \textbf{Optimizers:} Both Adam and SGD optimizers with default learning rates were explored.
    \item \textbf{Early Stopping:} Early stopping was enabled to mitigate overfitting and ensure convergence to optimal solutions.
\end{itemize}

\begin{figure}[h!]
    \includegraphics[width=1.02\textwidth]{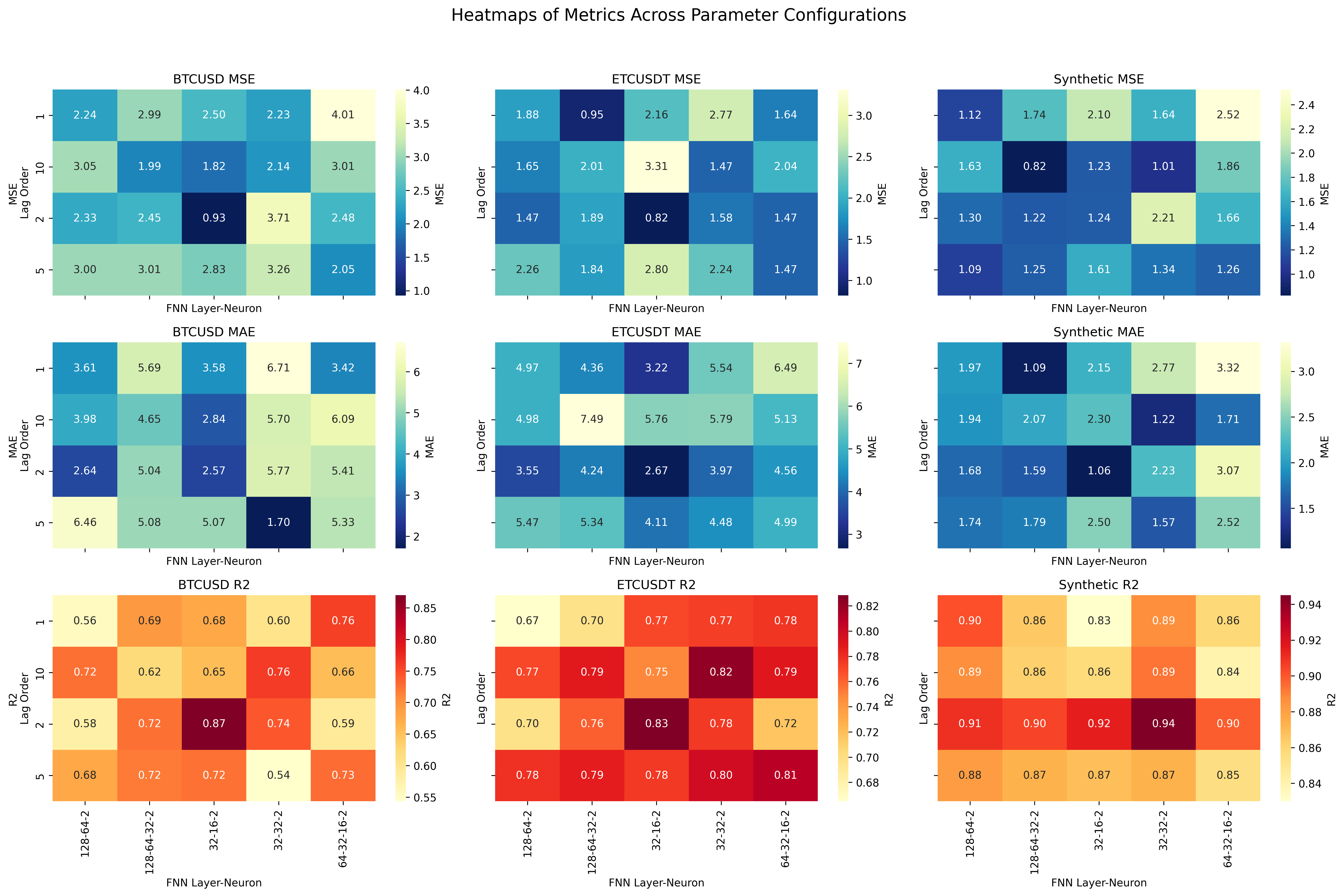}
    \caption{Sensitivity Analysis Heatmaps: Evaluation Metrics Across Parameter Configurations}
    \label{fig:sensitivity_heatmap}
\end{figure}

A total of 120 parameter combinations were evaluated manually and through optimization tools, including methods like Latin Hypercube Sampling (LHS) and Grid Search, to streamline the parameter selection process. These combinations spanned different neural layer architectures and lag orders, with the goal of identifying an optimal configuration. Model performance was assessed on three datasets using key evaluation metrics: Mean Squared Error (MSE), Mean Absolute Error (MAE), and \( R^2 \). The sensitivity analysis results, as shown in Figure~\ref{fig:sensitivity_heatmap}, provide a comparative view of configurations based on lag orders and FNN layers, highlighting the best-performing configuration. 

These optimization methods enabled the selection of parameters that achieved superior evaluation metrics compared to other configurations, underscoring the hybrid VAR-FNN model’s predictive power. For a comprehensive view of all configuration metrics, refer to the full results file available at ( Appendix~\ref{appendix:A5})

\textbf{Optimal Configuration:} The combination of lag order 2, FNN layer structure 32-16-2, ReLU activation function, and Adam optimizer provided the best results across all datasets, demonstrating low error rates and high accuracy metrics.

\subsubsection{Model Setup and Optimized Hyperparameters}
Both the FNN and Hybrid VAR-FNN models were constructed with identical neural network configurations in TensorFlow. The architecture and training parameters were defined as follows:
\begin{itemize}
    \item \textbf{Network Architecture}:
    \begin{itemize}
        \item \textbf{Layer 1}: 32 neurons, ReLU activation function.
        \item \textbf{Layer 2}: 16 neurons, ReLU activation function.
        \item \textbf{Output Layer for FNN Model}: Single neuron dedicated to OFI prediction.
        \item \textbf{Output Layer for Hybrid Model}: Two neurons for predicting buy and sell orders individually, from which OFI is calculated.
    \end{itemize}
    \item \textbf{Training Hyperparameters}:
    \begin{itemize}
        \item \textbf{Epochs}: 50, with early stopping to ensure model stability.
        \item \textbf{Batch Size}: 8, to optimize training time while maintaining accuracy.
        \item \textbf{Optimizer}: Adam, with a learning rate of 0.001.
    \end{itemize}
\end{itemize}

\subsection{Training Results}

The selected optimal configuration was used to train the hybrid VAR-FNN model on the primary dataset, achieving highly desirable training outcomes. Table~\ref{tab:training_results} presents the training metrics, showcasing that the hybrid model consistently achieved lower Mean Squared Error (MSE), Mean Absolute Error (MAE), and higher $R^2$ compared to the FNN-only model. This result demonstrates the hybrid model's enhanced ability to capture complex patterns in OFI dynamics, indicating strong predictive capabilities.
  
\begin{table}[h!]
\centering
\caption{Training Results for FNN and Hybrid VAR-FNN Models}
\label{tab:training_results}
\begin{tabular}{@{}lccc@{}}
\toprule
\textbf{Model}           & \textbf{MSE} & \textbf{MAE} & \textbf{$R^2$} \\ \midrule
FNN Only         & 0.00133          & 0.02240       & 0.9843       \\
Hybrid VAR-FNN   & 0.00127          & 0.00541       & 0.9946       \\ \bottomrule
\end{tabular}
\end{table}

\begin{figure}[h!]
    \includegraphics[width=1.0\textwidth]{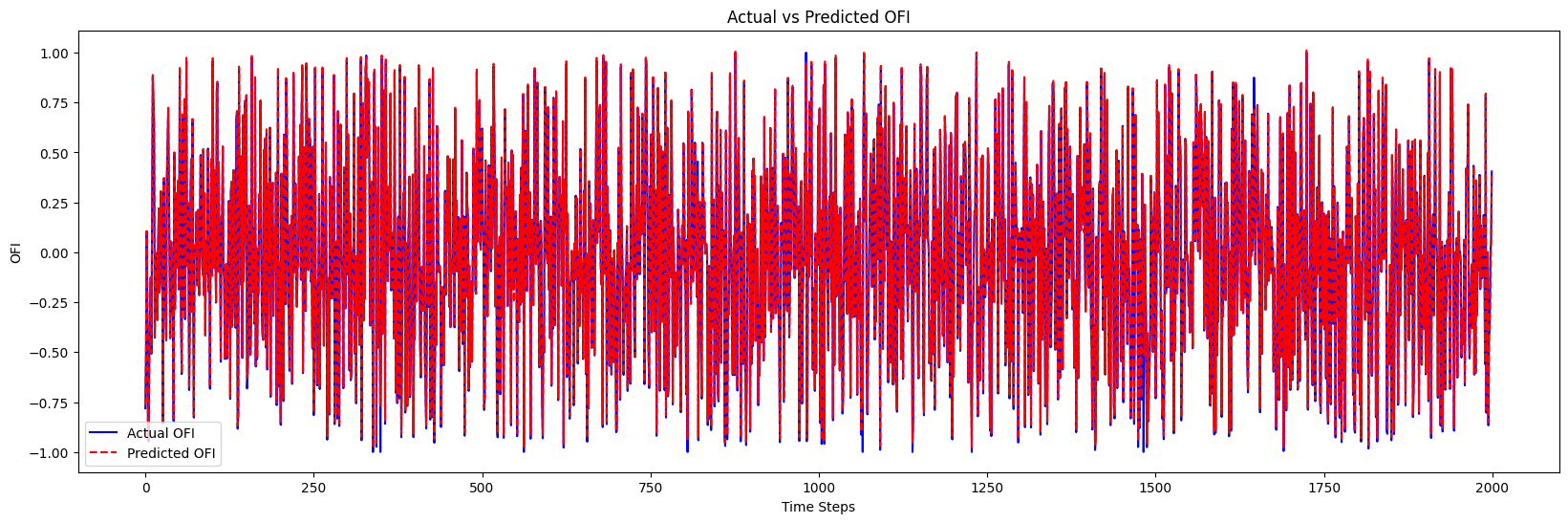}
    \caption{Predictions from the Hybrid VAR-FNN Model during Training.}
    \label{fig:train}
\end{figure}

\subsection{Validation Results}

Table~\ref{tab:validation_results} provides the model validation results across the three datasets. Metrics include MSE, MAE, $R^2$, trading intensity accuracy, and precision. The Hybrid VAR-FNN model consistently performs better than both the VAR-only and FNN-only models, particularly in accuracy and precision of trading intensity predictions.

\begin{table}[h!]
\centering
\caption{Validation Metrics for VAR, FNN, and Hybrid VAR-FNN Models}
\label{tab:validation_results}
\begin{tabular}{@{}p{2cm} p{3cm} p{1cm} p{1cm} p{1cm} p{1.8cm} p{1.8cm}@{}}
\toprule
\textbf{Dataset}        & \textbf{Model}       & \textbf{MSE} & \textbf{MAE} & \textbf{$R^2$} & \textbf{Accuracy (Intensity)} & \textbf{Precision (Intensity)} \\ \midrule
\addlinespace[0.2em]
BTCUSD           & VAR Only             & 0.675         & 0.757         & -0.002           & 46.61\%                          & 46.11\%                          \\
                        & FNN Only             & 0.021         & 0.078         & 0.970           & 97.43\%                          & 95.82\%                           \\
                        & Hybrid VAR-FNN       & 0.002         & 0.019        &   0.997 & 98.18\%                          & 96.32\%                          \\ \cmidrule[0.75pt]{1-7} 
\addlinespace[0.2em]
ETCUSDT          & VAR Only             & 0.856         & 0.881         & -0.165           & 67.06\%                         & 67.47\%                           \\
                        & FNN Only             & 0.107        & 0.019         & 0.974           & 95.29\%                          & 93.44\%                           \\
                        & Hybrid VAR-FNN       & 0.012         & 0.031         & 0.983           & 96.41\%                          & 95.01\%                           \\ \cmidrule[0.75pt]{1-7} 
\addlinespace[0.2em]
Synthetic          & VAR Only             & 0.228         & 0.387         & -0.001           & 50.37\%                          & 50.77\%                           \\
   Data                    & FNN Only             & 0.129        & 0.295         & 0.435           & 95.40\%                          & 98.77\%                               \\
                        
                        & Hybrid VAR-FNN       & 0.001         & 0.003         & 0.999           & 99.77\%                          & 99.54\%                           \\ \bottomrule
\end{tabular}
\end{table}

\subsection{Summary of Findings}

From the validation results, we observe the following key points:

\begin{itemize}
\item \textbf{Overall Performance}: The Hybrid VAR-FNN model consistently outperformed both the VAR and FNN-only models across all datasets. The MSE and MAE values for the Hybrid model were lower, and $R^2$ values were higher, indicating a more accurate fit.
    
    \item \textbf{Trading Intensity Prediction}: The Hybrid model also achieved superior accuracy and precision in predicting trading intensity signals, showing better differentiation between buy and sell pressure. On the BTCUSDT dataset, for example, the Hybrid model reached an accuracy of 98.18\%, compared to 97.43\% for the FNN-only model and 46.61\% for the VAR-only model.
    
    \item \textbf{Real vs. Synthetic Data}: The model performed exceptionally well on the synthetic dataset, achieving a high $R^2$ value of 0.999. This suggests that the Hybrid model can generalize effectively across different data types, with strong predictive capability in synthetic settings.
    
    \item \textbf{Improved Residual Modeling}: By capturing linear patterns through VAR and modeling non-linear residuals with FNN, the Hybrid VAR-FNN model demonstrated improved accuracy, particularly on volatile data from real markets. This dual approach allows for a more robust handling of market dynamics that are otherwise challenging for standalone models.
   \item \textbf{BTCUSD and ETHUSDT Analysis}: The results in Tables~\ref{tab:validation_results} and \ref{tab:BTCUSD_results} emphasize the accuracy of the Hybrid model in generating accurate trading signals and forecasting OFI. The table and visualizations highlight the predictive accuracy of the hybrid VAR-FNN model in forecasting OFI and generating accurate trading signals. The hybrid model consistently outperforms the standalone VAR and FNN models, as demonstrated by lower prediction errors and more accurate signal classifications.

    \item \textbf{Effectiveness of Hybrid Model}: The combined approach benefits from VAR’s ability to capture linear dependencies and the FNN’s capacity to learn non-linear residuals. This dual strategy enables the model to adapt to the volatility and complexities in high-frequency trading data, making it robust against market fluctuations.

    \item \textbf{Trading Signal Accuracy}: The hybrid model achieved high accuracy in predicting trading signals for both real and synthetic datasets, showing significant improvements over the VAR-only and FNN-only models. This enhanced predictive power can provide valuable insights for traders aiming to optimize their strategies based on OFI trends.

\end{itemize}

\subsubsection*{BTCUSD Dataset Results}

Table~\ref{tab:BTCUSD_results} provides the forecasted and actual OFI values, along with trading intensity signals for the BTCUSD dataset.

\begin{table}[h!]
\centering
\caption{BTCUSD Dataset Results}
\label{tab:BTCUSD_results}
\begin{tabular}{@{}p{0.8cm} p{1.2cm} p{1.5cm} p{1.5cm} p{1.5cm} p{1.5cm} p{1.2cm} p{1.2cm} p{1.2cm}@{}}
\toprule
\textbf{Index} & \textbf{Actual OFI} & \textbf{VAR Only OFI} & \textbf{FNN Model OFI} & \textbf{Hybrid Model OFI} & \textbf{Actual Signal} & \textbf{VAR Signal} & \textbf{FNN Signal} & \textbf{Hybrid Signal} \\ \midrule
0 & -0.3454 & 0.1135  & -0.9000 & -0.3315 & SELL & BUY & SELL & SELL \\
1 & -0.9430 & 0.1020  & -0.9000 & -0.9203 & SELL & BUY & SELL & SELL \\
2 & 0.2269  & 0.0065  & 0.6201  & 0.2256  & BUY  & BUY & BUY  & BUY  \\
3 & -0.1221 & 0.0067  & -0.5002 & -0.1203 & SELL & BUY & SELL & SELL \\
4 & 0.8429  & 0.0070  & 0.9000  & 0.8431  & BUY  & BUY & BUY  & BUY  \\
5 & -0.5616 & 0.0070  & -0.9000 & -0.5581 & SELL & BUY & SELL & SELL \\ \bottomrule
\end{tabular} 
\end{table}

\subsubsection*{ETHUSDT Dataset Visualization}

\begin{figure}[htbp]
\centering
  \begin{minipage}[t]{0.75\textwidth}
    \centering
    \includegraphics[width=\textwidth]{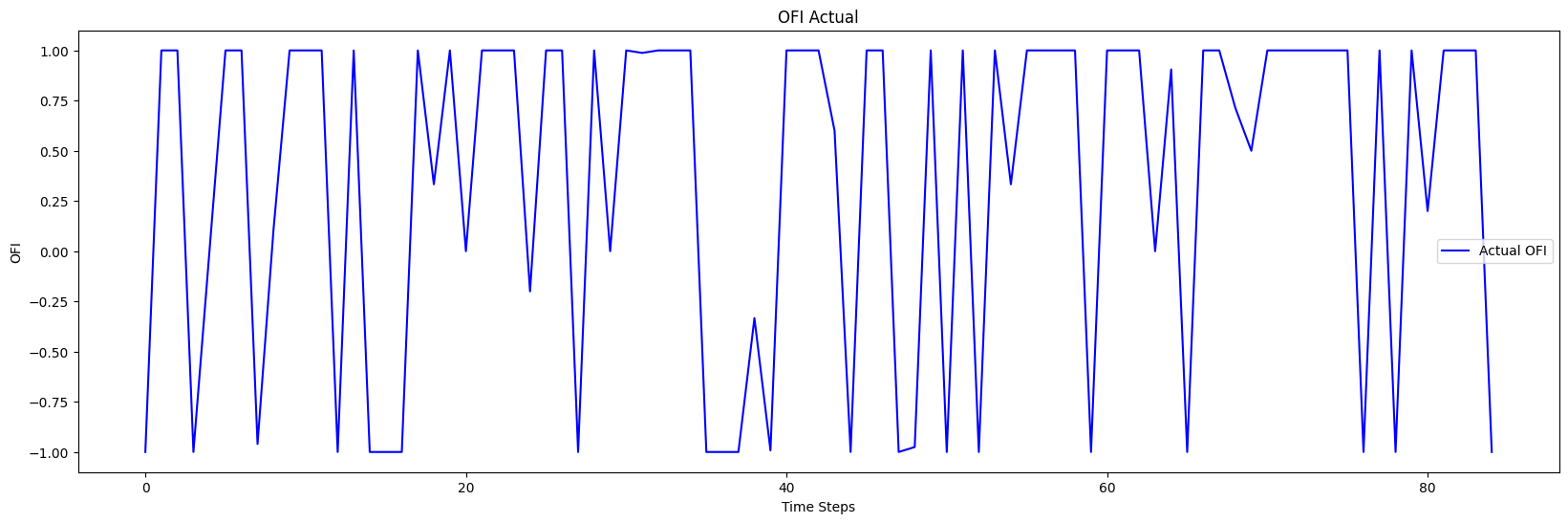}
    \caption{ETHUSDT Actual OFI}
    \label{fig:ethusdt_actual}
  \end{minipage}
  \hfill
  \begin{minipage}[t]{0.75\textwidth}
    \centering
    \includegraphics[width=\textwidth]{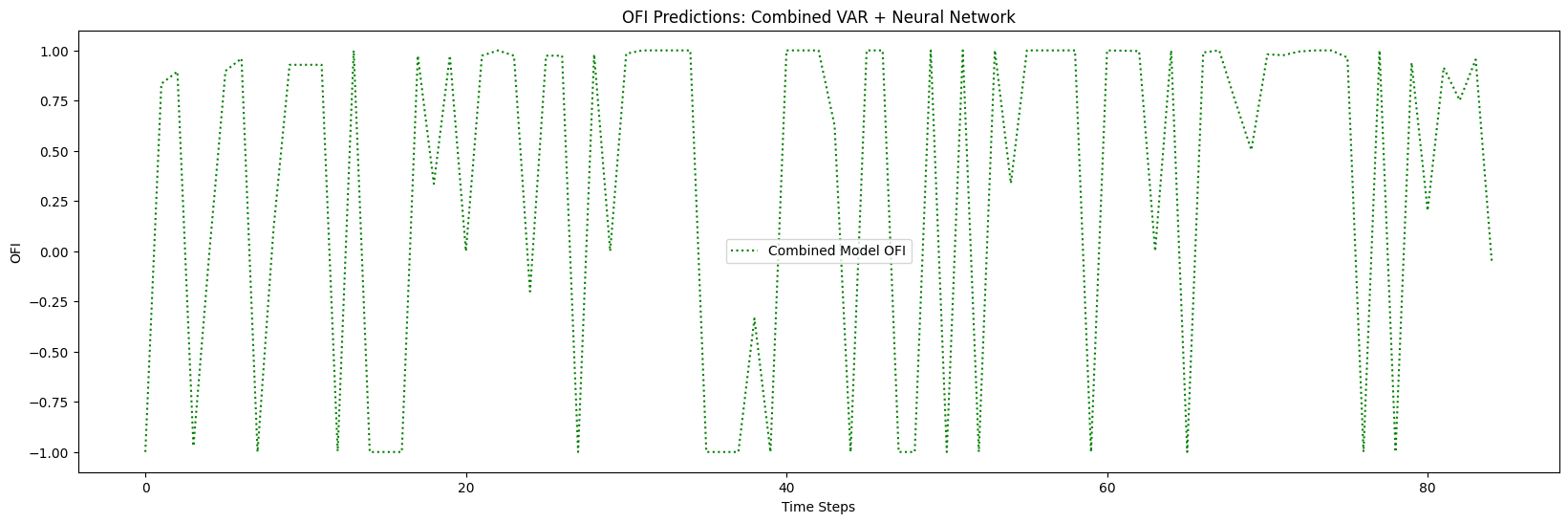}
    \caption{ETHUSDT Predicted OFI}
    \label{fig:ethusdt_predicted}
  \end{minipage}
  \hfill
  \begin{minipage}[t]{0.75\linewidth}
    \centering
    \includegraphics[width=\linewidth]{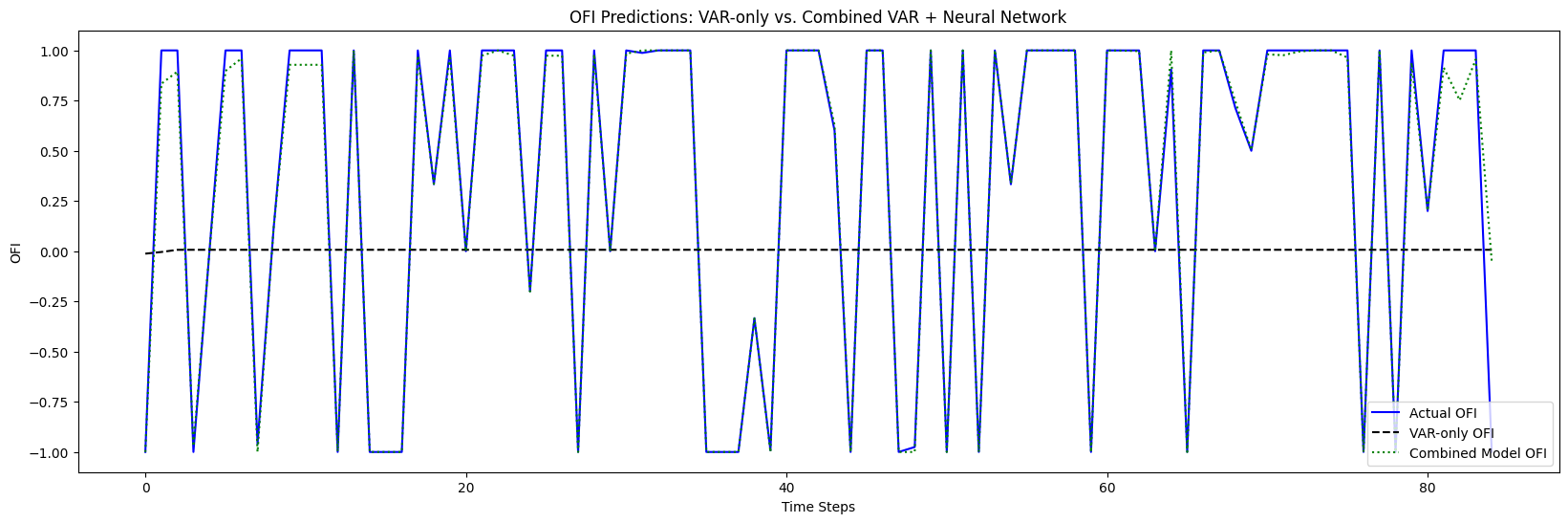}
    \caption{ETHUSDT Combined Model OFI Prediction}
    \label{fig:ethusdt_actual_predicted}
  \end{minipage}
\end{figure}

Figure~\ref{fig:ethusdt_actual}  , ~\ref{fig:ethusdt_predicted} and ~\ref{fig:ethusdt_actual_predicted}  illustrates the ETHUSDT dataset predictions - showcasing the actual OFI values, predicted OFI values from the model, and the final combined output.

\section{Conclusion, Limitations and Future Works}
\subsection{Conclusion}

This study presented a hybrid VAR-FNN model designed for Order Flow Imbalance (OFI) prediction in high-frequency trading contexts. By combining the strengths of the Vector Auto Regression (VAR) model to capture linear dependencies with a Feedforward Neural Network (FNN) for non-linear residuals, the hybrid model demonstrated superior performance compared to standalone VAR and FNN models. Through training on cryptocurrency data from Binance and validation on both real and synthetic datasets, the hybrid model achieved lower MSE and MAE, higher $R^2$ scores, and significantly improved trading intensity prediction accuracy.

Our findings indicate that the hybrid VAR-FNN model effectively leverages the interpretability of traditional econometric time series models with the adaptability of machine learning, making it a robust tool for predicting OFI in volatile, high-frequency environments. The model's ability to accurately predict directional trading signals offers valuable insights for market participants seeking to anticipate price movements and optimize trading strategies based on OFI trends.

\subsection{Limitations}

While the hybrid VAR-FNN model demonstrated notable improvements in predictive accuracy and signal precision, several limitations remain. The model's performance is heavily dependent on the quality and granularity of the high-frequency data, making it sensitive to any inaccuracies or noise in buy/sell order data. Additionally, this study focused on cryptocurrency data, specifically BTCUSD and ETHUSDT, so its generalizability across other asset classes (e.g., equities, forex) and varying market conditions is yet to be validated. Furthermore, the hybrid approach increases computational complexity, which might restrict its scalability for live trading applications that require real-time predictions. Finally, fixed hyperparameters and network architecture were effective for this study but may require further tuning and customization for broader applications or different datasets.

\subsection{Future Work}

To address these limitations and expand the potential of hybrid modeling approaches in high-frequency trading, several directions for future work are proposed. First, validating the hybrid VAR-FNN model across a diverse range of financial instruments, including equities, forex, and commodities, would provide insights into its adaptability and effectiveness across different asset classes and market conditions. Developing a streamlined, computationally efficient version of the model for real-time prediction could further enhance its applicability in live trading environments, potentially achieved by exploring model pruning or quantization. Incorporating advanced neural network architectures, such as Long Short-Term Memory (LSTM) networks or Graph Neural Networks (GNNs), may enhance the model’s ability to capture more complex temporal or relational patterns within trading data.

Additionally, employing automated hyperparameter tuning techniques, such as Bayesian Optimization, could help identify optimal configurations for various datasets and trading contexts, increasing the model’s robustness. Future research could also explore alternative hybrid models by integrating different econometric time series methods with machine learning models, such as combining ARIMA with recurrent neural networks or utilizing Universal Differential Equations (UDEs) to leverage both known dynamics and data-driven learning for OFI prediction. Lastly, incorporating additional relevant features, such as order book depth, trade volume, and historical volatility, may further improve the model's predictive power and provide a more comprehensive understanding of market dynamics.

In summary, the hybrid VAR-FNN model represents a promising approach for OFI prediction in high-frequency trading. By integrating econometric and machine learning techniques, this model offers a balanced solution that captures both linear and non-linear dynamics, making it a valuable tool for traders and researchers interested in enhancing the accuracy and robustness of OFI-based trading strategies.


\bibliographystyle{unsrtnat}
\bibliography{reference} 

\begin{thebibliography}{23}
\providecommand{\natexlab}[1]{#1}
\providecommand{\url}[1]{\texttt{#1}}
\expandafter\ifx\csname urlstyle\endcsname\relax
  \providecommand{\doi}[1]{doi: #1}\else
  \providecommand{\doi}{doi: \begingroup \urlstyle{rm}\Url}\fi

\bibitem[Cont et~al.(2023)Cont, Cucuringu, and Zhang]{Cont2023Cross-impactMarkets}
Rama Cont, Mihai Cucuringu, and Chao Zhang.
\newblock {Cross-impact of order flow imbalance in equity markets}.
\newblock \emph{Quantitative Finance}, 23\penalty0 (10):\penalty0 1373--1393, 10 2023.
\newblock ISSN 1469-7688.
\newblock \doi{10.1080/14697688.2023.2236159}.

\bibitem[Easley et~al.(2012)Easley, L{\'{o}}pez~de Prado, and O'Hara]{Easley2012FlowWorld}
David Easley, Marcos~M. L{\'{o}}pez~de Prado, and Maureen O'Hara.
\newblock {Flow Toxicity and Liquidity in a High-frequency World}.
\newblock \emph{Review of Financial Studies}, 25\penalty0 (5):\penalty0 1457--1493, 5 2012.
\newblock ISSN 0893-9454.
\newblock \doi{10.1093/rfs/hhs053}.

\bibitem[Kolm et~al.(2023)Kolm, Turiel, and Westray]{Kolm2023DeepBook}
Petter~N. Kolm, Jeremy Turiel, and Nicholas Westray.
\newblock {Deep order flow imbalance: Extracting alpha at multiple horizons from the limit order book}.
\newblock \emph{Mathematical Finance}, 33\penalty0 (4):\penalty0 1044--1081, 10 2023.
\newblock ISSN 0960-1627.
\newblock \doi{10.1111/mafi.12413}.

\bibitem[Murphy(2012)]{Murphy2012AMarkets}
Bernard Murphy, Finbarr.
\newblock {A vector-autoregression analysis of credit and liquidity factor dynamics in US LIBOR and Euribor swap markets}.
\newblock \emph{Journal of Economics and Finance}, 36\penalty0 (2):\penalty0 351--370, 4 2012.
\newblock ISSN 1055-0925.
\newblock \doi{10.1007/s12197-010-9122-2}.

\bibitem[Bacry et~al.(2016)Bacry, Jaisson, and Muzy]{Bacry2016EstimationDynamics}
Emmanuel Bacry, Thibault Jaisson, and Jean–François Muzy.
\newblock {Estimation of slowly decreasing Hawkes kernels: application to high-frequency order book dynamics}.
\newblock \emph{Quantitative Finance}, 16\penalty0 (8):\penalty0 1179--1201, 8 2016.
\newblock ISSN 14697696.
\newblock \doi{10.1080/14697688.2015.1123287}.

\bibitem[Huang et~al.(2007)Huang, LAI, NAKAMORI, WANG, and YU]{HUANG2007NEURALFORECASTING}
WEI Huang, KIN~KEUNG LAI, YOSHITERU NAKAMORI, SHOUYANG WANG, and LEAN YU.
\newblock {NEURAL NETWORKS IN FINANCE AND ECONOMICS FORECASTING}.
\newblock \emph{International Journal of Information Technology {\&} Decision Making}, 06\penalty0 (01):\penalty0 113--140, 3 2007.
\newblock ISSN 0219-6220.
\newblock \doi{10.1142/S021962200700237X}.

\bibitem[Maleki et~al.(2020)Maleki, Wraith, Mahmoudi, and Contreras-Reyes]{Maleki2020AsymmetricData}
Mohsen Maleki, Darren Wraith, Mohammad~R. Mahmoudi, and Javier~E. Contreras-Reyes.
\newblock {Asymmetric heavy-tailed vector auto-regressive processes with application to financial data}.
\newblock \emph{Journal of Statistical Computation and Simulation}, 90\penalty0 (2):\penalty0 324--340, 1 2020.
\newblock ISSN 0094-9655.
\newblock \doi{10.1080/00949655.2019.1680675}.

\bibitem[Smales(2013)]{Smales2013BondImbalance}
Lee~A. Smales.
\newblock {Bond futures and order imbalance}.
\newblock \emph{Journal of International Financial Markets, Institutions and Money}, 26:\penalty0 113--132, 10 2013.
\newblock ISSN 10424431.
\newblock \doi{10.1016/j.intfin.2013.05.006}.

\bibitem[Chan and Fong(2000)]{Chan2000TradeRelation}
Kalok Chan and Wai-Ming Fong.
\newblock {Trade size, order imbalance, and the volatility–volume relation}.
\newblock \emph{Journal of Financial Economics}, 57\penalty0 (2):\penalty0 247--273, 8 2000.
\newblock ISSN 0304405X.
\newblock \doi{10.1016/S0304-405X(00)00057-X}.

\bibitem[Cont et~al.(2014)Cont, Kukanov, and Stoikov]{Cont2014TheEvents}
R.~Cont, A.~Kukanov, and S.~Stoikov.
\newblock {The Price Impact of Order Book Events}.
\newblock \emph{Journal of Financial Econometrics}, 12\penalty0 (1):\penalty0 47--88, 1 2014.
\newblock ISSN 1479-8409.
\newblock \doi{10.1093/jjfinec/nbt003}.

\bibitem[Stock and Watson(2001)]{Stock2001VectorAutoregressions}
James~H Stock and Mark~W Watson.
\newblock {Vector Autoregressions}.
\newblock \emph{Journal of Economic Perspectives}, 15\penalty0 (4):\penalty0 101--115, 11 2001.
\newblock ISSN 0895-3309.
\newblock \doi{10.1257/jep.15.4.101}.

\bibitem[Toda and Phillips(1994)]{Toda1994VectorStudy}
Hiro~Y. Toda and Peter C.~B. Phillips.
\newblock {Vector autoregression and causality: a theoretical overview and simulation study}.
\newblock \emph{Econometric Reviews}, 13\penalty0 (2):\penalty0 259--285, 1 1994.
\newblock ISSN 0747-4938.
\newblock \doi{10.1080/07474939408800286}.

\bibitem[Anantha and Jain(2024)]{Anantha2024ForecastingImbalance}
Aditya~Nittur Anantha and Shashi Jain.
\newblock {Forecasting High Frequency Order Flow Imbalance}, 8 2024.
\newblock URL \url{https://arxiv.org/abs/2408.03594}.

\bibitem[Burrell and Folarin(1997)]{Burrell1997TheFinance}
P.~R. Burrell and B.~O. Folarin.
\newblock {The impact of neural networks in finance}.
\newblock \emph{Neural Computing {\&} Applications}, 6\penalty0 (4):\penalty0 193--200, 12 1997.
\newblock ISSN 0941-0643.
\newblock \doi{10.1007/BF01501506}.

\bibitem[Vellido(1999)]{Vellido1999Neural19921998}
A~Vellido.
\newblock {Neural networks in business: a survey of applications (1992–1998)}.
\newblock \emph{Expert Systems with Applications}, 17\penalty0 (1):\penalty0 51--70, 7 1999.
\newblock ISSN 09574174.
\newblock \doi{10.1016/S0957-4174(99)00016-0}.

\bibitem[Cartea et~al.(2018)Cartea, Donnelly, and Jaimungal]{Cartea2018EnhancingSignals}
Álvaro Cartea, Ryan Donnelly, and Sebastian Jaimungal.
\newblock {Enhancing trading strategies with order book signals}.
\newblock \emph{Applied Mathematical Finance}, 25\penalty0 (1):\penalty0 1--35, 1 2018.
\newblock ISSN 1350-486X.
\newblock \doi{10.1080/1350486X.2018.1434009}.

\bibitem[Bacry and Muzy(2014)]{Bacry2014HawkesDynamics}
Emmanuel Bacry and Jean-François Muzy.
\newblock {Hawkes model for price and trades high-frequency dynamics}.
\newblock \emph{Quantitative Finance}, 14\penalty0 (7):\penalty0 1147--1166, 7 2014.
\newblock ISSN 1469-7688.
\newblock \doi{10.1080/14697688.2014.897000}.

\bibitem[Wu and Siwasarit(2020)]{Wu2020CapturingKOSPI50}
Polin Wu and Wasin Siwasarit.
\newblock {Capturing the Order Imbalance with Hidden Markov Model: A Case of SET50 and KOSPI50}.
\newblock \emph{Asia-Pacific Financial Markets}, 27\penalty0 (1):\penalty0 115--144, 3 2020.
\newblock ISSN 1387-2834.
\newblock \doi{10.1007/s10690-019-09285-1}.

\bibitem[Bechler and Ludkovski(2015)]{Bechler2015OptimalImbalance}
Kyle Bechler and Michael Ludkovski.
\newblock {Optimal Execution with Dynamic Order Flow Imbalance}.
\newblock \emph{SIAM Journal on Financial Mathematics}, 6\penalty0 (1):\penalty0 1123--1151, 1 2015.
\newblock ISSN 1945-497X.
\newblock \doi{10.1137/140992254}.

\bibitem[Jiang(2011)]{Jiang2011OrderMarket}
Lei Jiang.
\newblock {Order Imbalance, Liquidity, and Market Efficiency: Evidence from the Chinese Stock Market}.
\newblock \emph{Managerial and Decision Economics}, 32\penalty0 (7):\penalty0 469--480, 10 2011.
\newblock ISSN 01436570.
\newblock \doi{10.1002/mde.1547}.

\bibitem[Fung and Yu(2007)]{Fung2007OrderPrices}
Joseph~K.W. Fung and Philip~L.H. Yu.
\newblock {Order imbalance and the dynamics of index and futures prices}.
\newblock \emph{Journal of Futures Markets}, 27\penalty0 (12):\penalty0 1129--1157, 12 2007.
\newblock ISSN 0270-7314.
\newblock \doi{10.1002/fut.20288}.

\bibitem[Kromkowski et~al.(2016)Kromkowski, Montgomery, Saha, Wu, and Beling]{Kromkowski2016EffectStates}
Andrew Kromkowski, Mason Montgomery, Kaustav Saha, Feiyin Wu, and Peter Beling.
\newblock {Effect of order flow imbalance on market impact across market states}.
\newblock In \emph{2016 IEEE Systems and Information Engineering Design Symposium (SIEDS)}, pages 298--302. IEEE, 4 2016.
\newblock ISBN 978-1-5090-0970-1.
\newblock \doi{10.1109/SIEDS.2016.7489318}.

\bibitem[Su et~al.(2012)Su, Huang, and Lin]{Su2012DynamicGainers}
Yong-Chern Su, Han-Ching Huang, and Shiue-Fang Lin.
\newblock {Dynamic relations between order imbalance, volatility and return of top gainers}.
\newblock \emph{Applied Economics}, 44\penalty0 (12):\penalty0 1509--1519, 4 2012.
\newblock ISSN 0003-6846.
\newblock \doi{10.1080/00036846.2010.543080}.

\end{thebibliography}


\appendix
\section{Appendix}

\subsection{VAR Model Training Summary}
\label{appendix:VAR_summary}

\begin{verbatim}
Summary of Regression Results
==================================
Model:                         VAR
Method:                        OLS
Date:           Thu, 31, Oct, 2024
Time:                     12:38:33
--------------------------------------------------------------------
No. of Equations:         2.00000    BIC:                    11.4739
Nobs:                     9998.00    HQIC:                   11.4692
Log likelihood:          -85685.2    FPE:                    95484.3
AIC:                      11.4667    Det(Omega_mle):         95388.9
--------------------------------------------------------------------
Results for equation buy_orders
=================================================================================
                    coefficient       std. error           t-stat            prob
---------------------------------------------------------------------------------
const                 29.357486         0.625923           46.903           0.000
L1.buy_orders          0.003770         0.010001            0.377           0.706
L1.sell_orders         0.011110         0.010021            1.109           0.268
L2.buy_orders         -0.000537         0.010001           -0.054           0.957
L2.sell_orders         0.016757         0.010022            1.672           0.095
=================================================================================
Results for equation sell_orders
=================================================================================
sell_orders     -0.000108     1.000000
---------------------------------------------------------------------------------
\end{verbatim}

\subsection{Time Complexity Analysis of FNN Modeling Algorithm}
\label{appendix:A2}
To analyze the time complexity of the FNN-only modeling algorithm, we break down each component:

\begin{itemize}
    \item \textbf{Forward and Backward Pass per Epoch}: For each epoch, the model performs a forward and backward pass through the neural network. Given:
        \begin{itemize}
            \item \( n \): The number of data points in the input.
            \item \( d \): The dimensionality of the input data.
            \item \( h \): The number of neurons in the hidden layers.
        \end{itemize}
        Each forward and backward pass has a time complexity of \( \mathcal{O}(n \cdot d \cdot h) \).
    
    \item \textbf{Total FNN Training Time over \( M \) Epochs}: With \( M \) epochs, the total time complexity is:
    \[
    \mathcal{O}(M \cdot n \cdot d \cdot h)
    \]
\end{itemize}

\textbf{Final Complexity Interpretation}

If \( M \), \( d \), and \( h \) are constants, the time complexity simplifies to:
\[
\mathcal{O}(n)
\]
indicating a linear time complexity with respect to \( n \) in this case. However, the full time complexity \( \mathcal{O}(M \cdot n \cdot d \cdot h) \) reflects the dependence on epochs, input dimensionality, and neurons in the hidden layers.

\subsection{Time Complexity Analysis of Hybrid VAR-FNN Model}
\label{appendix:A3}
To evaluate the time complexity of the Hybrid VAR-FNN modeling algorithm, we break down the computational cost for each component.

\subsubsection{Definitions and Assumptions}
\begin{itemize}
    \item \textbf{Input Size ($n$)}: Number of data points in the dataset.
    \item \textbf{VAR Model Complexity}:
        \begin{itemize}
            \item \textbf{Lag Order ($p$)}: The number of past observations in the VAR model.
        \end{itemize}
    \item \textbf{FNN Model Complexity}:
        \begin{itemize}
            \item \textbf{Epochs ($M$)}: Total training epochs.
            \item \textbf{Neurons per Layer ($h$)} and \textbf{Input Dimensionality ($d$)}: $h$ is the average number of neurons in each layer, and $d$ is the dimensionality of the input.
        \end{itemize}
\end{itemize}

\subsubsection{Step-by-Step Complexity Analysis}
\begin{enumerate}
    \item \textbf{VAR Model Training:}
    \begin{itemize}
        \item The VAR model uses Ordinary Least Squares (OLS) regression for each variable.
        \item For lag order $p$, complexity per variable is $\mathcal{O}(n \cdot p^2)$.
        \item With two variables (buy and sell orders), the total complexity becomes:
        \[
        \mathcal{O}(2 \cdot n \cdot p^2) = \mathcal{O}(n \cdot p^2)
        \]
    \end{itemize}
    
    \item \textbf{OFI Calculation:}
    \begin{itemize}
        \item Calculating OFI from VAR-predicted orders has complexity $\mathcal{O}(n)$, as it’s an element-wise operation.
    \end{itemize}
    
    \item \textbf{Residual Calculation:}
    \begin{itemize}
        \item Computing residuals by subtracting VAR predictions from actual orders is an element-wise operation with complexity $\mathcal{O}(n)$.
    \end{itemize}
    
    \item \textbf{FNN Training on Residuals:}
    \begin{itemize}
        \item For each epoch, the forward and backward passes through the FNN have complexity $\mathcal{O}(n \cdot d \cdot h)$.
        \item With $M$ epochs, the total complexity for FNN training becomes:
        \[
        \mathcal{O}(M \cdot n \cdot d \cdot h)
        \]
    \end{itemize}
    
    \item \textbf{Combining VAR and FNN-Predicted OFI:}
    \begin{itemize}
        \item The final OFI calculation requires element-wise addition, with a complexity of $\mathcal{O}(n)$.
    \end{itemize}
\end{enumerate}

\subsubsection{Total Complexity}

The combined time complexity is:
\[
\mathcal{O}(n \cdot p^2) + \mathcal{O}(n) + \mathcal{O}(n) + \mathcal{O}(M \cdot n \cdot d \cdot h) + \mathcal{O}(n)
\]
Simplifying, the dominant terms give:
\[
\mathcal{O}(n \cdot p^2 + M \cdot n \cdot d \cdot h)
\]
In cases where $p$, $M$, $d$, and $h$ are constants, the time complexity reduces to:
\[
\mathcal{O}(n)
\]
However, generally, the model’s complexity is approximately linear $\mathcal{O}(n \cdot (p^2 + M \cdot d \cdot h))$, reflecting both the VAR and FNN components.

\subsection{Experiment Logs and Training Curves}

To provide transparency on the model's training stability, the following figures illustrate the training and validation loss curves for both the FNN-only model and the hybrid VAR-FNN model. These plots demonstrate the convergence behavior over epochs and help validate the model’s generalization capability.

\begin{figure}[htbp]
  \begin{minipage}[t]{0.48\textwidth}
  \centering
  \includegraphics[width=\textwidth]{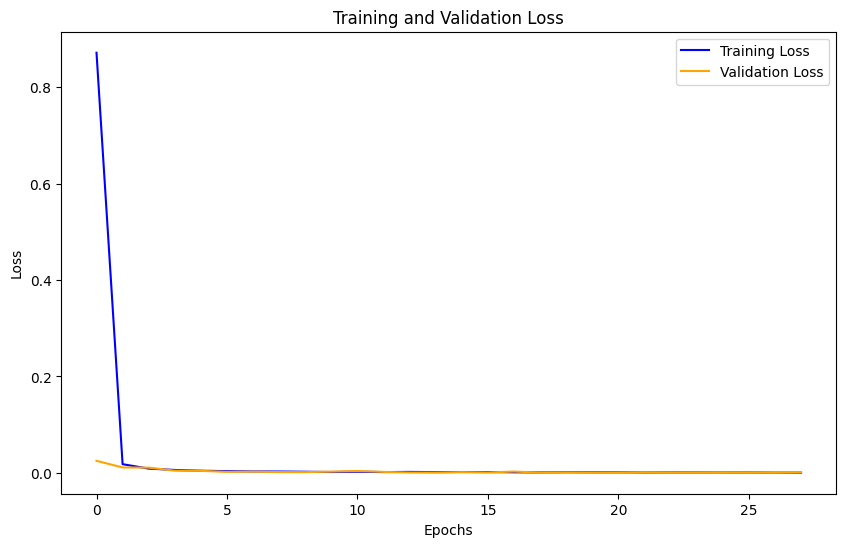} 
  \caption{ FNN-Only Model}
  \label{fig:fnn_training_curve}
  \end{minipage}
  \hfill
  \begin{minipage}[t]{0.48\textwidth}
  \centering
  \includegraphics[width=\textwidth]{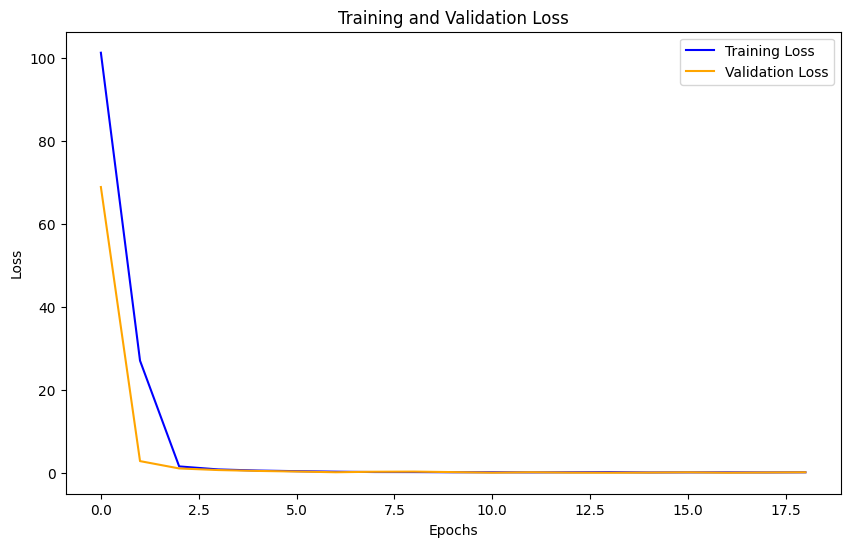} 
  \caption{Hybrid VAR-FNN Model}
  \label{fig:var_fnn_training_curve}
  \end{minipage}
\end{figure}

1. \textbf{Training and Validation Loss Curves for FNN-Only Model} \\
    Figure~\ref{fig:fnn_training_curve} shows the loss curve for the standalone FNN model. The graph indicates how the model's loss decreased over the epochs, along with validation loss to monitor overfitting or underfitting behavior.

2. \textbf{Training and Validation Loss Curves for Hybrid VAR-FNN Model} \\
    Figure~\ref{fig:var_fnn_training_curve} presents the loss curve for the hybrid VAR-FNN model. The steady decline in loss, along with minimal divergence between training and validation loss, suggests effective learning and good generalization on unseen data.

\subsection{GitHub Repository Link}
\label{appendix:A5}
For the complete source code, including data preprocessing, model training, and evaluation scripts, please refer to the GitHub repository: \href{https://github.com/Shafiq-Abdu/OFI_Main.git}{GitHub Repository Link}.

\begin{itemize}
    \item \textbf{Instructions}: Clone the repository and follow the instructions in the \texttt{README.md} file to set up the environment and run the models.
    \item \textbf{Dependencies}: All required dependencies are listed in the \texttt{requirements.txt} file.
    \item \textbf{Execution}: Detailed instructions on training and testing the models are provided in the repository documentation.
    \item \textbf{Sensitivity Analysis : } Comprehensive view of all configuration metrics and full result available as  \href{https://github.com/Shafiq-Abdu/OFI_Main/blob/main/heatmap/sensitivity_analysis_results.csv}{csv\_file}.
\end{itemize}

\end{document}